\documentstyle[12pt]{article}
\catcode`\@=11 \@addtoreset{equation}{section}  
   
\title{Quantum Theory Requires Gravity and \\
Superrelativity }
\author{P. Leifer}
\date{ School of Physics and Astronomy \\ Raymond and 
Beverly Sackler Faculty of Exact Sciences \\
Mortimer and Raymond Sackler Institute of Advanced Studies \\ Tel-Aviv University, Tel-Aviv 69978, Israel \\
e-mail:leifer@ccsg.tau.ac.il}
\begin{document}
\maketitle
\begin{abstract}
The ordinary quantum theory points out that general relativity is 
negligible for spatial distances up to the Planck scale $l_P=(\hbar G/c^3)^{1/2} \sim 10^{-33} cm$. Consistency in the foundations of the quantum theory requires a``soft'' spacetime structure of the GR at essentially longer length. However, for some reasons this appears to be not enough.
A new framework (``superrelativity'') for the desirable generalization of the foundation of quantum theory is proposed. A generalized non-linear Klein-Gordon equation  has been derived in order to shape a stable wave packet.
 
\end{abstract} 

\section{Introduction}
In both special relativity (SR) and general relativity
(GR) we deal with the 
motion of a system of material points.
In particular, in GR, one can think of a description
in terms of the local tangent space (freely falling
frame) at every point. From this point of view, the 
dynamical variable, constructed in each tangent space,
depend on the state of the system (the spacetime 
neighbourhood). But in the framework of quantum theory
notions of both the ``material point'' and the ``freely falling frame'' should be replaced. Namely, notions
of ``quantum state'' and ``local functional frame'' 
are some natural generalization of main classical elements.
If one takes into account this type of structure in the  
the framework of quantum theory, we see that its 
manifistation in GR is merely a spacial case of this 
more general quantum structure - ``Superrelativity''. 

All attempts to achieve a reasonable generalization of the 
quantum theory which are based on the Einstein's principle of GR lead to major difficulties. The reason is that in the framework of GR 
only the spacetime structure has been modified, but not  
the quantum 
state space. That is one treats quantum particles as 
a material point in a spacetime and internal degrees of freedom are 
expressed by the evolution of second quantized amplitudes of a probability 
in the state space. This point of view came from Dirac's classical 
articles [1,2]. In his work Ref. [2], the introduction of internal degrees of 
freedom for an electron (spin) is based on the principle of special 
relativity. Dirac treats the ``doubling'' of the number of states 
as an evidence of the non-locality of the electron. However, he formulates 
the local problem to find a Hamiltonian linear in  the operator of momentum $\hat{p}_\mu$ for the point charged electron. But one suspects that 
besides the spin degree of freedom, there are some ``hidden degrees of freedom'' which describe a spatial distribution
of both charge and mass. 
Therefore, I think we should \it modify \rm two of Schr\"odinger's original ideas: 

1. Wave description of elementary particles as wave packets [3],

2. Wave propagation in the configuration space [4,5].

We propose here a simple possibility for a ``shaping'' of 
a stable wave 
packet (``droplet'') from solutions of the linear  Klein-Gordon equation by the action of geodesic flow in the Hilbert projective space $CP(N-1)$ [6-9].

In principle, it is impossible to distinguish ``external'' and ''internal'' 
degrees of freedom of a quantum system. \it Therefore, 
to my 
mind, one must take into account \rm \bf an 
entanglement \rm \it of  
the spacetime and the state space 
of a quantum particle\rm. Supergravity realizes it in so-called superspace. Instead, I propose to describe quantum 
particles in a single \bf projective Hilbert space \rm $CP(N-1)$.
\it It is a quite different approach to the non-distinguishability 
of quantum systems and the decoherence of 
isolated ``bits''. In this approach spacetime arises as an auxiliary 
entity for a description of  motion of quantum 
integrals \rm. Moreover, in an attempt to unify of fundamental interactions we try to achieve this just 
in the 
$CP(N-1)$. That is, a distance in this space is \bf the square root of 
an action \rm and not spacetime interval [7-9]. One 
should find a new geometric structure with this \it 
interval of 
action \rm for an imbedding of both internal and 
external kinds of symmetry. 

$SU(N)$ symmetry is an example of so-called internal symmetry of 
elementary particles. This symmetry is broken up to the isotropy group $H=U(1)\otimes U(N-1)$ of the pure quantum state. We will show 
that a character of this break-down is connected to the geometric structure mentioned 
above as the coset $SU(N)/S[U(1) \otimes 
U(N-1)]$ is connected with the projective Hilbert space CP(N-1). This 
geometry is associated with the spacetime distribution of physical 
carriers of charges, masses etc., i.e. elementary 
particles.
  
\section{Goal, Price and Method}
{\bf Goal} We wish to create a concentrated 
self-interacting field configuration ``droplet'' which has 
its own surrounding gauge field (intrinsic potential). It should play the role of the model of non-local quantum 
particles
in the framework of the causal approach to quantum theory
[6-9,11]. 

Now, at the 70th birthday of Quantum Mechanics, it is 
clear 
that the price of success must be very high. 
Nevertheless we ought to pay, if we want to reach a reasonable comprehension of the quantum entity.

{\bf Price} We should forget about the spacetime priority.
Quantum state space and states themselves are fundamental
elements of the ``quantum essentiality''.
That is our physical {\it dynamical} spacetime 
arises as a geometry of moving nonlocal but
concentrated quantum particles. Therefore we have to find
some geometry reflecting both quantum   
features and the possibility of the quasiclassical 
approximation in a natural way. 

As a matter of fact this point of view is 
neither shocking nor novel. For example, Y. Ne'eman wrote
that the ``quantum reality'' of complex quantum amplitudes
is ``represented by Hilbert space, rather than by 
spacetime''[10]. But ordinary Hilbert space is closely
related to ordinary spacetime and it must be clear
how to unify them. One of the ways is well known now. This is supersymmetry and its local version - supergravity. But this method of unification
acts as if both spacetime and the space of internal degrees 
of fredom themselves were independent entities [11].
 
{\bf Method} I propose a new geometric framework in order to unify both
 ``external'' and ``internal'' degrees
of freedom which I will call ``Superrelativity''(SuperR). {\it The
SuperR physically means that there is a unified 
self-interacting physical field of 
de Broglie-Schr\"odinger-Bohm which is associated with the geometry
of the projective Hilbert space CP(N-1). This space has a constant positive holomorphic sectional curvature.
The curvature gives an intrinsic interaction of modes of extended quantum particles. Since the projective Hilbert space
CP(N-1) is a homogeneous manifold, there are both local and global conservation laws. 
Hereof there arise so-called local (in CP(N-1)) dynamical variables which depend on the state of a quantum system (local coordinates); dynamical variables in  special or in general relativity  depend on the state of a motion of the material point as well}.

\section{Principle of ``Superrelativity''}
I try to achieve of the `peaceful
coexistence' between principles of Einstein's relativity
(locality and determinism) and quantum theory which is
incompatible with local supplementary parameters. Note,
this should be done in the framework of a {\it non-linear approach}.
This requires a deep reconstruction of the relationships
between spacetime structure and the quantum state space
(in my model I used the complex projective Hilbert space of indefinite dimensionality $CP(\infty)$ or, in some approximation, the complex projective Hilbert space $CP(N-1)$ of  finite dimension). Briefly speaking, I tried to expand
a wave description on `quantum particles' (in the spirit of
de Broglie-Schr\"odinger-Bohm [12]) in the framework of a
more wide and ``soft'' structure than spacetime, even
than the curved Einstein spacetime of general relativity. The
physical meaning of this structure (from the mathematical point of view it is the tangent fiber bundle over $CP(\infty)$ or $CP(N-1)$) may be expressed in a principle
of ``Superrelativity''. 

In order to realize this principle we should introduce 
two fundamental notions. They are ``state'' and
``transition'' of a quantum system. These notions should be
used instead of ``material point'' and ``event'' in
special relativity (SR) or in GR. Note that the spacetime
structure ``grows'' from
the geometric structure  of the state space. That is
at the microlevel spacetime coordinates (in Einstein's
sense) have a merely interpolating sense (in the reference
Minkowski spacetime) at best and do not exist at worst.

Superrelativity is  based on  simple physical facts:

1. {\it Every event in the sense of special or general  relativity is a quantum transition}. Therefore, one cannot invoke {\it only} our spacetime experience and should deal with a state space. 
For instance, in an arbitrary chosen point $B$ of the Minkowski spacetime there
does not exist a natural notion of the ``same direction'' relative to the initial direction in the initial point $A$. At every point one can define the ``direction'' as the direction of some physical vector field. Therefore, a comparison of directions must be reduced to the comparison of the vector fields.  As a matter of fact one should establish a law of  ``parallel transport'' of quantum dynamical variables of the vector fields configuration. This law takes place in the state space 
of the field configuration and not in spacetime itself.
It is a generalization of the well known Pancharatnam problem of the comparison of polarizations of two beams
[11]. 

2. {\it The structure of the state space reflects 
symmetries of real quantum particles}. 
$SU(N)$ symmetry is an example of so-called internal symmetry of 
elementary particles. Here we seek only the simplest  possibility,
assuming that this symmetry is broken up to the isotropy group $H=U(1) \otimes U(N-1)$ of the pure quantum state. 

We assume that the content of the ``superrelativity principle'' is as follow:
{\it The general unitary motion of pure quantum states may be locally reduced to
geodesic motion in the projective Hilbert space by 
introduction of some gauge (compensation) field for self-
preservation of the geodesic field configuration (droplet). Herein
a surrounding field arises; we try to identify it with
a ``physical field'' in the ordinary spacetime}. Note, that ``superrelativity'' assumes some 
``superequivalence'' of unified physical field and geometric
properties of the base manifold (in our model it is projective Hilbert space which is equipped with the 
generalized Fubini-Study metric [6-9,11]). It is
useful to compare the equivalence principle of Einstein
and ``superequivalence'' principle.

The Einstein's equivalence principle has often been subjected to criticism. It is correct that it is fulfilled only
locally. It is correct that the  absence of a gravitation field at a point implies a zero value of the Riemann tensor of the spacetime  and this condition does not depend
on the character of an observer's motion. There exists an
opinion, however, that this principle should be discarded. If we understand this principle {\it literally} as the equivalence
of curvature of the spacetime (gravity) and the arbitrary
``physical fields'' as a reason for the accelerated motion,
then perhaps this really should be done. But I think that we ought
to take into account the {\it general aspiration} of Einstein.
I mean  that he tried to build a {\it unified field 
theory}. Everyone knows it must be a quantum theory. 
In the framework of this quantum  theory all fields 
have a unified nature and 
therefore {\it a novel equivalence principle in Einstein's
spirit} should concern {\it quantum states}, not
material points. Such notions as ``accelerated motion''
and ``uniform motion'' are no longer applicable to quantum
states. Hence, one must invent {\it some new classification
of motion of quantum states} (i.e. classification of quantum
transitions). Then we should formulate a ``superequivalence''
principle on the base of this classification. In our case
this classification will be based upon a geometrically invariant distinction of local and global conservation laws in the 
projective Hilbert state space. In accordance with this 
intrinsic classification, we have two kinds of motions:

A.Unitary ``rotations'' of the ``ellipsoid of 
polarization'' under transformations from the isotropy 
group $U(1)\otimes U(N-1)$ of the pure quantum state
(Higgs modes).

B.Geodesic ``motion'' of pure quantum states in the projective Hilbert space as a
hidden (virtual) transition under transformations from the 
coset $SU(N)/S[U(1)\otimes U(N-1)]$. They are pure ``deformations''
of the ``ellipsoid of polarization'' (Goldstone modes). 

Pure quantum states of ``isolated'' quantum systems are rays 
$\{|\Psi>\}=A\exp{i\alpha}\sum_{a=0}^{N-1} \Psi^a|a,x>$
and they
belong to the projective Hilbert space $CP(N-1)$ [13]. There are appropriate local coordinates $\pi^i_{(b)}$ of the chart atlas 
$U_b= \{|\Psi>=\sum _{a=0}^{N-1}\Psi^a|a,x>:\Psi^b \not=0 \}$ in CP(N-1). For $b=0$ one has
\begin{equation}
\pi^i_{(0)}=\Psi^i/\Psi^0,       
\end{equation}
where $1\leq i \leq N-1$. 
Then the fundamental tensor of Fubini-Study metric in $CP(N)$ is 
\begin{equation}
g_{ik*}=2\hbar\frac{(1 + \sum_{s=1}^{N-1}|\pi^s|^2) \delta_{ik} 
-\pi^{i*} \pi^k}{(1+\sum_{s=1}^{N-1}|\pi^s|^2)^2}. 
\end{equation}
[13-15].  At first sight the superposition principle 
permits one  to work with the relative amplitudes and relative phases, i.e., one must forget about modulus of 
the wave function. However perhaps the superposition principle serves
merely as a very good approximation. This is a quite 
natural
assumption if we try to build a {\it nonlinear quantum
theory where this principle, of course, does not act}.
Then the modulus has a physical meaning and, hence, 
should be taken into account. I shall assume that 
projective
symmetry is broken up to the symmetry of the K\"ahler 
manifold with metric (3.3). It may be done by using 
the generalized Fubini-Study metric tensor $G_{ik*}$ in CP(N-1) [6-9,11], which is defined by the formula
\begin{equation}
G_{ik*}=2 \hbar R^2 {(R^2 + \sum_{s=1}^{N-1}|\pi^s|^2) \delta_{ik} 
-\pi^{i*} \pi^k \over (R^2+\sum_{s=1}^{N-1}|\pi^s|^2)^2}. 
\end{equation}
The real part of (3.3) is a Riemannian structure and  the imaginary part is a symplectic one. In addition, the natural connection (see below) determines an intrinsic gauge potential. The symplectic structure plays an important role in the geometric phase. The Riemannian structure and curvature of $CP(N-1)$ is
closely connected with the density of Schr\"odinge's wave, because  
the {\it generalized Fubini-Study metric} (3.3) can then be regarded as an induced Riemann metric of $CP(N-1)$. It is obtained by the ``stereographic projection'' of rays of the Hilbert space C(N) from the ``density sphere'' $\sum _{a=0}^{N-1} |\Psi^a|^2 = R^2$  with radius $R$. Therefore the
well known geometric interpretation of Planck's constant
is appropriate as a normalizing factor for the radius of the
sphere $S^{N^2-2}$ in the $AlgSU(N)$ [15] for average $<A>=\frac{<\Psi|\hat{A}|\Psi>}
{<\Psi|\Psi>}$. But internal (Riemann) geometry takes place
on the ``density sphere'' $S^{2N-1}$ in the Hilbert space. 
That is we have a {\it spectral parameter $R$ for the ``foliation'' of the tangent bundle over CP(N-1) where the unified fundamental interaction acts}. As a matter of fact it is  the lift of a ``trace of the quantum transition'' in the base CP(N-1) into the tangent fiber bundle of the``experimental environment of external fields''. Then the semiclassical limit may be achieved if $R \to \infty$ [8]. We will show (see below) that this limit physically may be achieved very easily for  a {\it finite} value of $R$. Hence, I think it more natural to identify
the curvature of state space not with Planck's constant
but with fine structure constant $c=1/R^2=\alpha$. 
In our case Planck's constant is merely a normalizing factor as well as in Ref. [15].  We, therefore, do 
not treat the ``semiclassical limit'' in terms of the limit as Planck's constant $\hbar$ tends to zero, but as $R$ tends to infinity. That is we can avoid the conclusion that ``radius of holomorphic sectional curvature goes to zero as $\hbar \to 0$'' which is very paradoxical [15]. We have
here an example of a different kind of non-linearity 
then in Weinberg approach [16,17].

\section{Generalized Pancharatnam connection}
The generalized Pancharatnam's problem of 
comparison 
of the phases of beams is akin to the Shapere-Wilczek approach 
to the comparison of shapes of 
deformable bodies in their gauge kinematics [18].
In our case the projective Hilbert space 
$CP(N-1)$
 takes the place of the space of some ``unlocated 
shapes'' [11]. Of course, it is impossible to understand 
the 
``shape of wave packet'' - droplet,- literally as the 
shape of the quantum particle in the real 
spacetime. 
The droplet is a geodesic (periodic) deformation of Fourier 
components of the initial solution of 
Klein-Gordon 
equation under trasformations from the coset 
$SU(N)/S[U(1) \otimes U(N-1)]$. That is the
problem 
which arises in our case and should be stated as 
followes: {\it  what are the dynamical variables that
correspond to sequence of deformations of solution 
of the initial 
linear equation?} Here $CP(N-1)$ is a base 
manifold and $U(1) \otimes U(N-1)$ is the structure
group in a fiber.
 
The problem of the comparison of real spatial shapes, which undergo large deformations, has not yet been solved [18]. 
But in our case the ``instantaneous shape of a wave packet'' is  represented by {\it known vector fields of polarizations as functions 
of relative Fourier components themselves} and deformations of this 
``shape of ellipsoid of polarization'' lay in the coset 
$SU(N)/S[U(1) \otimes U(N-1)]$ [6-9]. 
That is the natural connection in $CP(N-1)$ 
\begin{equation}
\Gamma^i_{kl} = -2 \frac{\delta^i_k \pi^{l*} + \delta^i_l \pi^{k*}}
{R^2 + \sum_s^{N-1} |\pi^s|^2} 
\end{equation}
which corresponds to the Fubini-Study metric (3.3), can help us to compare these shapes. Perhaps the most obvious example of the same kind is a representation
of the polarization states of photons or electrons by points of the Poincar\'e
sphere [19]. The ``shape'' in this case is indeed thee shape of the ellipse of polarization. The ellipse conserves its own shape along every ``parallel
of latitude'' but the orientation of this ellipse is smoothly  and periodically 
changing.
We will show that {\it the connection (4.1) determines a quite natural intrinsic
gauge potential of a local frame rotation in a tangent space of $CP(N-1)$
and, therefore, renormalization
of dynamical variables. Relationships between Goldsone's and Higgs's
modes arise in an absolutely natural way also. Namely,
the shape of the graphic (4.1) is similar to the well known
artificial potential surface 
$V=\lambda^2|\pi|^4-\mu^2|\pi|^2$ for the 
illustration of the spontaneously broken symmetry,
however our ``potential'' (4.1) is finite anywhere}.

\section{Quantum ``Droplet''} 
The general form of Newton's second law is indifferent to the
type of force. Only development of electromagnetic theory, 
that is a particular kind of force, leds to the relativistic generalization 
of the classical mechanics. The Schr\"odinger equation of ordinary 
quantum mechanics is indifferent to the choice of  a potential also.
However, at short distances this potential can not be arbitrary. 
Therefore, it is not enough to use the Hertz metric of configuration space,  which contains an arbitrary potential
of the ``environment'' [4,5]. Underlying ``hidden'' degrees of 
freedom, connected with internal symmetries of elementary particles, should be used for ``shaping'' an intrinsic potential which is a ``particle'' itself. This approach coinsides in general with de Broglie's idea [20]. I will build our model in the spirit of two main approaches:

1.Schr\"odinger's method of coherent states for stable
wave packet ``shaping'' [3],

2.Bohm approach to the nonlinear origin of fundamental
equations of elementary particles [12].
 
{\it The essential new element of our approach is the 
action of geodesic flow in the configuration space on the relative Fourier components $CP(N-1)$. Since the principle of least action arises in 
$CP(N-1)$ as a principle of least curvature, geodesic  flow  in $CP(N-1)$
plays an important role [8,9,11]. Its integral curves (geodesics) are stable and closed (periodic)}. 

It should to be clear that we are not ready yet to present
a quite consistent theory. Our aim is to show how
we can move toward this desirable goal. Then we can
introduce the notion of a ``reference Minkowski spacetime''
as an analogy of a ``screen 2-space'' on our PC. The mouse
has really 7 degrees of freedom in the original 3-space, but the pointer has only 2 degrees of freedom. I think in the quantum ``reality'' we have the 
same situation. Namely, the quantum field system (mouse) can have in general an indefinite number of degrees of freedom, but the ``centrum of mass''
of the  droplet,  which takes the place of the ``pointer of quantum
transition under registration'', has only 4-spacetime
degrees of freedom. The connection between the ``quantum mouse''
and ``pointer'' may be realized by some field model in spirit of Bohm [12].

As an example consider the effective self-interaction scalar field $\Phi$ in
``reference Minkowski spacetime'' [11]. That is we neglect  any 
dynamical effects (like effects of the spacetime curvature in general
relativity) in {\it this manifold}.  The physical spacetime should arise as 
a geometry of moving droplet where these dynamical effects may be observed.
We wish to write a Lagrangian for non-linear interaction, which is Poincar\'e invariant, and, therefore, choose a field which depends 
only on the ``radial'' variable $\rho$ , i.e.
$\Phi = \Phi (\rho)$, where $\rho^2 =  x_\mu x^\mu$ and $x^\mu$
corresponds to a relative spacetime coordinate (emerging, for example,
from some underlying dynamical model for self-interaction). Since 
$\Phi_{,\mu} = \frac{\partial \Phi}{ \partial x^ \mu}=
\frac {x_ \mu}{\rho} \frac{d \Phi}{d \rho}$ and
$\Phi^{,\mu} = \frac {\partial \Phi}{ \partial x_ \mu}=
\frac {x^ \mu}{\rho} \frac{d \Phi}{d \rho} $, a Lagrangian density may be written as
\begin{equation}
\cal L\rm=\frac{1}{2} \Phi^*_\mu \Phi^\mu - U(\Phi(\rho)) = 
\frac{1}{2} \left|\frac{d\Phi}{d\rho} \right|^2 -  U(\Phi(\rho)),
\end{equation}
where we have assumed a general form for the effective self-interaction
term $U(\Phi(\rho))$.The equation of motion of the scalar field
acquires the form of the ordinary differential (nonlinear in general)  equation
\begin{equation}
\frac{d^2 \Phi^*}{d \rho^2} + (3/\rho) \frac{ d \Phi^*}{d \rho} +
2\frac{\partial U(\Phi(\rho))}{\partial \Phi} =0 \ \ \ \ \ \  c.c.
\end{equation}
If the potential $U(\Phi(\rho))$ has the form 
$U(\Phi(\rho))=
\frac{1}{2}(mc/\hbar)^2\Phi^* \Phi =\frac{1}{2}\alpha^2 |\Phi|^2$,
then (5.2) is the linear Lommel equation 
\begin{equation}
\frac{d^2 \Phi^*}{d \rho^2} + (3/\rho) \frac{ d \Phi^*}{d \rho} + \alpha^2 \Phi^* = 0, \ \ \ \ \ \ c.c.
\end{equation}
for which a solution is the Bessel function [21]
$\Phi = \rho^{-1} J_{-1} (\alpha \rho)$.
We should note that in the timelike sector of the spacetime where
$\rho^{'2}=-\rho^2 >0$ corresponds to the de Broglie equations
which has a solution which is the modified Bessel functions
$\chi=\rho^{'-1} I_{-1} (\alpha \rho^{'}$). 
However  a ``deformation'' of  these solutions into solutions of some 
{\it effectively nonlinear Klein-Gordon or de Broglie equation} by the geodesic flow is interesting
for our purpose. This point is a crucial difference 
between
our approach and, say, a model of Rubakov-Saha [22].  I wish to emphasize the connection of our model with the so-called ``off-shell models'' [23].

If we choose the the classical radius of the electron $r_0=\frac{e^2}{mc^2}$
as the unit of distance scale $\rho = x r_0$, 
then $(mc/\hbar)^2$ in (5.3) becomes the fine structure constant 
$\alpha = \frac{e^2}{\hbar c}$.  Let's suppose $y=(\rho/r_0)^2$.

It is well known that on the interval $(-\infty,\infty)$ the set of orthogonal Hermitian functions 
$|n,y>=\phi_n(y)=(2^n n! \sqrt\pi)^{-1/2}\exp(-y^2 /2)
H_n(y)$
is complete and one can represent 
a solution of (5.3) in the $y$-variable as a Fourier series
\begin{equation}
|\Phi> = x^{-1} J_{-1} (\alpha x)=\sum_{k=0}^{\infty} \Phi^k \phi_k(x^2)= \sum_{k=0}^{\infty} \Phi^k |k,y>,
\end{equation}
where 
\begin{equation}
\Phi^m=\frac{-1}{2^m m! \sqrt\pi}\int_{-\infty}^{\infty}
dy \exp(-\frac{y^2}{2}) H_m(y)y^{-1/2} J_1(\alpha y^{1/2})
\end{equation}
Our geodesic flow acts on these Fourier components. It is convenient to transform the state covector (5.5)
to the ``vacuum'' form \\
$\Phi_0=exp(i \omega) ||\Phi||(1,0,...,0,...)$ by a set of matrices $\hat{G}$, obtained as follows [6-9]. It is easily seen that for a vector of this form, geodesic flow is generated by a general linear combination of ``creation-annihilation'' operators 
\begin{equation}
\hat{B}= \left(
\matrix{
0&f^{1*}&f^{2*}&.&.&.&f^{N-1*} \cr
f^1&0&0&.&.&.&0 \cr
f^2&0&0&.&.&.&0 \cr
.&.&.&. &.&.&. \cr
.&.&.&. &.&.&. \cr
.&.&.&. &.&.&. \cr
f^{N-1}&0&0&.&.&.&0
}
\right ). 
\end{equation}
The flow is then given by the unitary matrix
$\hat{T}(\tau,g)=\exp(i\tau\hat{B})=$
\begin{equation}
\left(
\matrix{
\cos\Theta&\frac{-f^{1*}}{g} \sin\Theta&.&.&.&\frac{-f^{N-1*}}{g}\sin\Theta \cr
\frac{f^1}{g}\sin\Theta&1+[\frac{|f^1|}{g}]^2 (\cos\Theta -1)&.&.&.& \frac{f^1 f^{N-1*}}{g^2}(\cos\Theta-1) \cr
.&.&.&.&.&.\cr
.&.&.&.&.&.\cr
.&.&.&.&.&.\cr
\frac{f^{N-1}}{g}\sin\Theta&\frac{f^{1*} f^{N-1}}{g^2}
(\cos\Theta-1)&.&.&.&1+[\frac{|f^{N-1}|}{g}]^2 (\cos\Theta -1)
}
\right),
\end{equation}
where $g=\sqrt{\sum_{k=1}^{N-1}|f^k|^2},\Theta=g\tau$ 
[6,9,11].
The form of the periodic geodesic ``deformation'' of the  initial solution of Eq.(5.3) is represented by the formula 
\begin{equation}
|\Psi (\tau,g,x)> = \sum^{\infty}_{m,n=0} |n,x> \Phi^m [\hat{G} \hat{T}(\tau,g) 
\hat{G}^{-1}]^n_m.
\end{equation}
That is, we have geodesic ``generation''
of nonlocal droplet with finite action. We try 
to build a quantum dynamical variables over Fourier modes 
of some
field like in Ref. [24] in distinction from Schr\"odinger's
wave function of coordinates of material points [4,5]. 
But instead of a cavity massless scalar field of Ref. [24] we used the massive ``Lorentz-radial'' scalar field $\Phi(x)$.

The uniform rotation (5.7) of  a state vector in the Hilbert space should be connected with a motion of the local coordinate (3.1) (Goldstone modes). On the other hand every geodesic could be rigidly
transformed from one to another by transformations from the isotropy group $U(1) \otimes U(N-1)$ of the ``vacuum'' state, because CP(N-1) is a ``totally geodesic manifold'' [14] (Higgs modes). That is one can identify any geodesic as a ``rigid framework'' for the shape of the stable wave packet - droplet. They look like ``closed strings'' in $CP(N-1)$.
In particular cases it may be transformed into a geodesic in $CP(1)$. Then $\Pi=R \exp(i\alpha) \tan l$ is a solution of the 
equation of a geodesic in CP(1)
\begin{equation}
\frac{d^2 \Pi}{dl^2} - \frac{2 \Pi^*}{R^2+|\Pi|^2} (\frac{d\Pi}{dl})^2=0.
\end{equation}
In the general case CP(N-1) we have $\pi^i(\lambda) = R(f^i/g) 
\tan \Theta$ for the \it uniform rotation \rm  of the ``vacuum'' state 
$\Phi_0$ in the original Hilbert space $C^N$ with
the rate g. However, \it relative to the natural measure in $CP(N-1)$, i.e. relative to the length $l(\pi,\pi^*)$ of a curve (action), this ``rotation'' is far from uniform. \rm
The relationship between these rates may be expressed by following equation
\begin{equation}
\frac{d^2 \Theta}{dl^2} + 2[1+2/R](\frac{d\Theta}{dl})^2
\tan \Theta=0.
\end{equation}
We can present a numerical solution of this equation as a 
dependence of the uniform rotation parameter $\tau$ on the natural (canonical) parameter $l$.
It may be shown that $\lim_{l,R \to \infty}\Theta(l,R)=\pi/2$.
{\it That is orthogonality of pure states for large action
may be achieved just in the ``semiclassical limit''              [8].Therefore the geometric structure (curvature)
of the Hilbert projective space has an essential physical meaning in terms of decoherence}.

In order to define the surrounding field of our scalar carrier-droplet we should introduce  new important notions.  

\section{Local Dynamical Variables and Field Equation} 
The problem of building of consistent quantum dynamical
variables (time and frequency) using the underlying
symmetries of quantum fields was raised by M.-T.Jaekel
and S.Reynaund  [25]. The main aim of this work is to 
clarify the notion of some kinds of spacetime 
transformation in a framework of `a novel conception of spacetime which would be free from its difficulties inherited from classical physics'. It is an absolutely 
legitimate question but it seems to me to require a
generalization. In comparison with Ref.[25] there are two differences in our approach to a similar problem: the 
first one is our conception of the {\it quantum transition} instead of ``event'' of SR or GR and CP(N-1) construction
as fundamental physical structure; the second one is a
4-dimensional spacetime structure instead of 
2-dimensional model.
\newtheorem{guess}{Definition}
\begin{guess}
Local (state-dependent) dynamical variables are tangent vector fields to the $CP(N-1)$ associated with one-
parameter subgroups of unitary transformations [6-9].
\end{guess} 
That is, we now refer to the term "local" as a fact of a dependence on the 
coordinates (3.1) in the $CP(N-1)$ as in Ref. [26].
We should find the relationship between the linear
representations of $SU(N)$ group by an  
``polarization operator'' $\hat{P} \in AlgSU(N)$ which 
does not depend on the state of the quantum system and the  nonlinear representation (realization) of the group symmetry in which the infinitesimal operator of the transformation depends on the
state.  In the linear representation of the action of $SU(N)$
we have
\begin{equation}
|\Psi(\epsilon)>=\exp(-i\epsilon \hat{P}) |\Psi>.
\end{equation}
For a full description of a group dynamics by pure quantum states, we shall  
use coherent states in $CP(N-1)$. Let's assume $\hat{P}_\sigma$ is
one of the $1\leq \sigma \leq N^2-1$ directions in the 
group manifold.
Then 
\begin{equation}
D_\sigma(\hat{P})=\Phi^i_\sigma (\pi,P)\frac{\delta}{\delta \pi^i}
+\Phi^{i*}_\sigma (\pi,P)\frac{\delta}{\delta \pi^{i*}},
\end{equation}
where
\begin{equation}
\Phi_{\sigma}^i(\pi;P_{\sigma}) = \lim_{\epsilon \to 0} \epsilon^{-1}
\biggl\{\frac{[\exp(i\epsilon P_{\sigma})]_m^i \Psi^m}{[\exp(i \epsilon P_{\sigma}]_m^k
\Psi^m }-\frac{\Psi^i}{\Psi^k} \biggr\}=
 \lim_{\epsilon \to 0} \epsilon^{-1} \{ \pi^i(\epsilon P_{\sigma}) -\pi^i \}
\end{equation}
are the local (in $CP(N-1)$) state-dependent components of generators 
of the $SU(N)$ group, which are studied in [6-9].
So in the general 
case for group transformations of more than one parameter we have 
a vector field for 
the group action on $CP(N-1)$ by some set of dynamical variables
$\hat{P}_1,...,\hat{P}_\sigma ,..., \hat{P}_{N^2-1}$, as
\begin{equation}
V_{P}(\pi,\pi^*)=\sum_\sigma  [\Phi^i_\sigma (\pi,P)\frac{\delta}{\delta \pi^i}
+\Phi^{i*}_\sigma (\pi,P)\frac{\delta}{\delta \pi^{i*}}] \epsilon^{\sigma}.
\end{equation}
Then the differential of some differentiable function $F(\pi,\pi^*)$ is
\begin{equation}
\delta_{P} F(\pi,\pi^*)=D_\sigma(\hat{P})F(\pi,\pi^*) \epsilon^\sigma,
\end{equation}
and, in particular, we have
\begin{equation}
\delta_{P}\pi^i=\Phi^i_\sigma (\pi,P) \epsilon^{\sigma},
\delta_{P}\pi^{i*}=\Phi^{i*}_\sigma (\pi,P) \epsilon^{\sigma}.
\end{equation}
For example, realizing rotations $\hat{s}_x,\hat{s}_y,
\hat{s}_z$ from 
$Alg SU(2)$, one has  
\begin{eqnarray} 
D_x(s)=-\frac{\hbar}{2}[[1-\pi^2]\frac{\delta}{\delta \pi}-
[1-\pi^{*2}]\frac{\delta}{\delta \pi^*}], \cr
D_y(s)=\frac{\hbar}{2}[[1+\pi^2]\frac{\delta}{\delta \pi}+
[1+\pi^{*2}]\frac{\delta}{\delta \pi^*}],\cr
D_z(s)=\hbar[-\pi \frac{\delta}{\delta \pi}+\pi^* \frac{\delta}{\delta \pi^*}]. 
\end{eqnarray}
Then, we have well known commutation relations
\begin{equation}
[D_\mu(s),D_\nu(s)]_-=- i \hbar \epsilon_{\mu \nu \sigma} D_\sigma
(s).
\end{equation}
For a three-level system, the realization of a dynamical $SU(3)$
group symmetry is provided by an 8-dimensional local vector field
[6], where $\hat{\lambda}_1,...,\hat{\lambda}_8$ are
the Gell-Mann matrices, i.e.
\begin{eqnarray}
D_1(\lambda)=i \frac{\hbar}{2}[[-1+(\pi^1)^2]\frac{\delta}{\delta \pi^1}
+\pi^1 \pi^2 \frac{\delta}{\delta \pi^2}
+[-1+(\pi^{1*})^2]\frac{\delta}{\delta \pi^{1*}}
+\pi^{1*} \pi^{2*} \frac{\delta}{\delta \pi^{2*}}] ,  \cr
D_2(\lambda)=i \frac{\hbar}{2}[[1+(\pi^1)^2]\frac{\delta}{\delta \pi^1}
+\pi^1 \pi^2 \frac{\delta}{\delta \pi^2}
-[1+(\pi^{1*})^2]\frac{\delta}{\delta \pi^{1*}}
-\pi^{1*} \pi^{2*} \frac{\delta}{\delta \pi^{2*}}] , \cr
D_3(\lambda)=-\frac{\hbar}{2}[\pi^2 \frac{\delta}{\delta \pi^2}
+\pi^{2*} \frac{\delta}{\delta \pi^{2*}}], \cr
D_4(\lambda)= \frac{\hbar}{2}[[-1+(\pi^2)^2]\frac{\delta}{\delta \pi^2}
+\pi^1 \pi^2 \frac{\delta}{\delta \pi^1}
+[-1+(\pi^{2*})^2]\frac{\delta}{\delta \pi^{2*}}
+\pi^{1*} \pi^{2*} \frac{\delta}{\delta \pi^{1*}}] , \cr
D_5(\lambda)= \frac{\hbar}{2}[[1+(\pi^2)^2]\frac{\delta}{\delta \pi^2}
+\pi^1 \pi^2 \frac{\delta}{\delta \pi^1}
-[1+(\pi^{2*})^2]\frac{\delta}{\delta \pi^{2*}}
-\pi^{1*} \pi^{2*} \frac{\delta}{\delta \pi^{1*}}], \cr
D_6(\lambda)=-\frac{\hbar}{2}[\pi^2 \frac{\delta}{\delta \pi^1}
+\pi^1 \frac{\delta}{\delta \pi^2}
-\pi^{2*}\frac{\delta}{\delta \pi^{1*}}
-\pi^{1*} \frac{\delta}{\delta \pi^{2*}}] , \cr
D_7(\lambda)=-\frac{\hbar}{2}[\pi^2 \frac{\delta}{\delta \pi^1}
-\pi^1 \frac{\delta}{\delta \pi^2}
-\pi^{2*}\frac{\delta}{\delta \pi^{1*}}
+\pi^{1*} \frac{\delta}{\delta \pi^{2*}}] , \cr
D_8(\lambda)=3\hbar[\pi^2 \frac{\delta}{\delta \pi^2}
-\pi^{2*} \frac{\delta}{\delta \pi^{2*}}].  
\end{eqnarray}

In each of $N$ charts of the local coordinates (3.1) these vector fields 
might be distinguished to two parts :
Goldstone subspace $B$ and Higgs subspace $H$ with 
commutation relations of $Z_2$ -graded algebra $Alg SU(N$):
 $[\hat{H},\hat{H}]_- \subset {H}, [\hat{H},\hat{B}]_- \subset {B}, [\hat{B},\hat{B}]_- \subset {H}$ 
like ordinary (state-independent) elements of  $Alg SU(N)$ [6-9]. 
 
In order to establish relationship 
between ``internal'' parameters of the droplet  and ``external'' propagation of the 
scalar field near the light cone in the 
``reference spacetime'', we should ``lift'' a geodesic cyclic virtual transition in $CP(\infty)$ (as a 
model of a single particle) into the fiber bundle. 

Namely, if we assume that {\it in accordance with the
``superequivalence principle'' an infinitesimal geodesic ``shift'' of dynamical variables could be compensated by an infinitesimal transformations of the basis in Hilbert space, then one can get some effective self-interaction potential as an addition to the mass 
term in original Klein-Gordon equation in the Lommel's form (5.3)}.  
We will label hereafter vectors of the Hilbert space by Dirac's notations
$|...>$ and tangent vectors to $CP(N-1)$ or  $CP(\infty)$ by 
arrows over letters, $\vec \xi$, for example. Then one has a definition
of the rate of a state vector  changing
$|v(x)>=-(i/\hbar)\hat{P}|\Psi (x)>$.
Of course, any dynamical variable
of the scalar field, charge, for example, defines a rate of change of the state vector. For us it is interesting now 
to consider an ``evolution'' during the quantum transition  along geodesic between vacuum state $|\Phi_0>$ and the state vector (5.4).
This ``evolution'' corresponds to a fast ``proper time''
$\tau$ [23] which is associated with frequency which should be close to the meson mass if one want to have a spatial propagation of the droplet close to the classical radius of electron.

The ``descent'' of the vector field $|v(x)>$ onto the base manifold 
$CP(\infty)$ is a mapping by the formula
\begin{eqnarray}
f_{*(\Psi^0,..., \Psi^m,...)} |v(x)> 
=\frac{d}{d \tau}(\frac{\Psi^1}{\Psi^0},..., \frac{\Psi^i}{\Psi^0},...)\Bigl|_0 \cr
=-(i/\hbar)[P^1_0+(P^1_k-P^0_k \pi^1)\pi^k,..., P^i_0+(P^i_k-P^0_k \pi^i)\pi^k,...] \cr 
= \vec \xi \in T_\pi CP(\infty).
\end{eqnarray}
If (and only if) one starts from the ``vacuum'' state in an arbitrary direction in $CP(\infty)$, i.e. from zeroth local
coordinates $\pi^1=...=\pi^i=...=0$ one has
\begin{equation}
f_{*(1,0,...,0,...)} |v_0>=-(i/\hbar)[P^1_0,...,P^i_0,...]
\end{equation}
and, therefore, one can identify $P^i_0=f^i$ or $\hat{P}
=\hat{B}$. In the general case  this is not correct. In order to find concrete values of $f^i$ for the ``evolution'' of the vacuum
state toward our solution, we must
use the formula $\pi^i(\tau) = R(f^i/g)\tan \Theta$.
If state $<\Phi|$ is not so far from $<\Phi_0|=(1,0,0,...,0)$
we can span them by an unique geodesic:  
\begin{equation}
g^{-1}(1,0,0,...,0)\hat{T}(\tau,g)=
R^{-1}(\Phi^0,\Phi^1,...,\Phi^N).
\end{equation}
Then one has $cos\Theta=|\Phi^0|/R$, $|f^i|= g|\Phi^i|
(R^2-|\Phi^0|^2)^{-1/2}$ and $\arg f^i=\arg \Phi^i$
(up to the general phase). Thus
we know f-elements from (5.6) for the transformation of the vacuum vector into the solution of the Lommel equation. The transformation of this solution into the vacuum vector is induced by elements of matrix of the general ``polarization
operator''
\begin{equation}
\hat{P}=\hat{G}^{-1}(\Phi)\hat{B}(\Phi)\hat{G}(\Phi) 
\end{equation}
Note, that complicated form of the matrix $\hat{P}$ is
the consequence of the fact that {\it subgroup} $H=U(1)\otimes U(N-1)$
{\it is not the normal (invariant) subgroup of the group $SU(N)$}. This operator determines a tangent vector field
$\vec \xi$ (6.10). At a point  
$\pi+\delta \pi$ in $CP(\infty)$ the ``shifted'' field  
\begin{equation}
\vec \xi+ \delta \vec \xi=\vec \xi+\frac{\delta \vec \xi}{\delta l} \delta l
\end{equation}
contains the derivative $\frac{\delta \vec \xi}{\delta l}$, which is not, in the
general case, a tangent vector to $CP(\infty)$, but the 
{\it covariant derivative}
\begin{equation}
\frac{\Delta \xi^i}{\delta l}=\frac{\delta \xi^i}{\delta l}+\Gamma^i_{km}
\xi^k \frac{\delta \pi^m}{\delta l}, \ \ \ \ \ \ c.c.
\end{equation}
is a tangent vector to $CP(\infty)$. Now we should ``lift'' the new tangent
vector $\xi^i + \Delta \xi^i$ into the original Hilbert space $\cal H \rm$,
that is, one needs to realize two mappings: $f^{-1}:CP(\infty) \to \cal H \rm $
\begin{eqnarray}
f^{-1}(\pi^1+\Delta \pi^1,...,\pi^i + \Delta \pi^i,...) \cr
=[\Psi^0,\Psi^0 (\pi^1+\Delta \pi^1),...,\Psi^0(\pi^i + \Delta \pi^i),...] \cr
=[\Psi^0,\Psi^1+\Psi^0 \Delta \pi^1),...,\Psi^i+\Psi^0 \Delta \pi^i),...]
\end{eqnarray}
and then
\begin{eqnarray}
f^{-1}_{* \pi+\delta \pi}(\vec \xi + \Delta \vec \xi)
=\frac{d}{d\tau}[\Psi^0,\Psi^0 (\pi^1+\Delta \pi^1),..., 
\Psi^0(\pi^i + \Delta \pi^i),...]\Big|_0 \cr
=[v^0, v^1+ \frac{d}{d\tau}(\Psi^0 \Delta \pi^1)\Big|_0,..., 
 v^i+\frac{d}{d\tau}(\Psi^0 \Delta \pi^i)\Big|_0,...]. \end{eqnarray}

Herein a non-parallel (in the general case) local vector field, corresponding to some local dynamical variable like (6.2), arises along our geodesic. Our aim is to find the
total field mass, charge etc. In 
order to do this we should use the parallel transport of the dynamical variables [27,6-9,11].

It may be shown in our original Hilbert space $\cal H \rm$ 
that the term $|dv>$ arises as an additional rate of a change of some general state vector $|\Psi>$
\begin{eqnarray}
|dv>=-(i/\hbar)d\hat{P}|\Psi> \cr
= [0, \frac{d}{d\tau}(\Psi^0 \Delta \pi^1)\Bigl|_0|1,x>,...,\frac{d}{d\tau}(\Psi^0 \Delta \pi^i)\Bigl|_0|i,x>,...] , 
\end{eqnarray}
where $\Delta \pi^i=-\Gamma^i_{km}\xi^k d\pi^m \tau$.
Then $<\Psi|d\hat{P}|\Psi>$ may be  treated as an ``instantaneous'' 
self-interacting potential of the scalar droplet associated with 
the infinitesimal gauge transformation of the local frame  
with coefficients (4.1). {\it That is self-preservation of the droplet (unperturbed geodesic sequence of virtual 
transitions) may be achieved by the radiation
of the gauge (compensation) field due to the  renormalization of dynamical variables  and ``rotation''
of the ellipsoid of polarization}.

In order to find the additional terms to the
Lagrangian density (5.1) induced by infinitesimal 
gauge transformations of the local frame in the tangent
space to $CP(N-1)$ (6.16), one should take into account
the fact that Fourier components in (5.5) do not depend on
spacetime coordinates in the case of the ``Lorentz-radial''
symmetry. That is, only spacetime derivatives  of the 
basis Hermitian functions 
\begin{eqnarray}
\nabla|n,y>= -\frac{2\vec{x}\exp(-y^2 /2)}{r_0^2\sqrt{
2^n n! \sqrt\pi}}(yH_n(y)-2nH_{n-1}(y)),\cr 
\frac{\partial |n,y>}{\partial t}=
\frac{2c^2 t\exp(-y^2 /2)}{r_0^2\sqrt{
2^n n! \sqrt\pi}}(yH_n(y)-2nH_{n-1}(y)) 
\end{eqnarray}
arise in the formula for Lagrangian density which is
induced by the ``geodesic variation'' of the inital Lagrangian
(5.1). On the other hand only Fourier components (5.5)  are subjected to the variation by the geodesic flow.
The state vector (6.14) inherits the geomeric structure
of the $CP(\infty)$ and perturbed Lagrangian as follows:
\begin{eqnarray}
\cal L'\rm=\cal L\rm(\Psi+\Delta \Psi)=(\Psi+\Delta \Psi)^{m*}\frac{\partial <m,y|}{\partial t}
\frac{\partial |n,y>}{\partial t}(\Psi+\Delta \Psi)^{n}\cr
-(\Psi+\Delta \Psi)^{m*}\nabla <m,y|\nabla |n,y>(\Psi+\Delta \Psi)^n \cr
-\alpha^2 (\Psi+\Delta \Psi)^{m*}<m,y|n,y>(\Psi+\Delta \Psi)^n,
\end{eqnarray}
where $\Delta \Psi^i=- \Psi^0 \Gamma^i_{km}\xi^k d\pi^m \tau$.

It is useful to compare the new Lagrangian with the 
well known Lagrangians of both abelian 
\begin{eqnarray}
\cal L_A\rm=-\frac{1}{4}F_{\mu \nu} F^{\mu \nu}+
\frac{1}{2}(\partial-ieA_{\mu})\psi^*(\partial-ieA^{\mu})\psi-\frac{1}{4}\lambda(|\psi|^2-F^2)^2
\end{eqnarray}
and non-abelian 
\begin{eqnarray}
\cal L_{NA}\rm=-\frac{1}{4}G^a_{\mu \nu} G^{a \mu \nu}+
\frac{1}{2}(D_{\mu}\psi^{*a})(D^{\mu}\psi^a)-
\frac{1}{4}\lambda(\psi^{*a}\psi_a-F^2)^2
\end{eqnarray}
Higgs models. Here $F_{\mu \nu}=\partial_{\mu}A_{\nu}-
\partial_{\nu}A_{\mu}$ and $G^a_{\mu \nu}=\partial_{\mu}A^a_{\nu}-
\partial_{\nu}A^a_{\mu}+g' \epsilon_{abc}A^b_{\mu}A^c_{\nu}$
are field tensors, $D_{\mu}\psi^a=\partial_{\mu}\psi^a+
g' \epsilon_{abc}A^b_{\mu}\psi^c$ is the so-called ``covariant
derivative'' and $F$ is the ``modulus of the vacuum''. There are some important differences between
our Lagrangian (6.20) and Lagrangians (6.21),(6.22):

1.First of all we have a single fundamental self-interacting scalar field $\Phi$ and modes of this field correspond  to
the energy of one of the N topological vacuums (the choice of the vacuum is, as a matter of fact, the choice of the 
chart of the base manifold, the projective Hilbert space 
$CP(N-1)$).

2.Instead of three parameters ($F,\lambda, g'$) of the models (6.21),(6.22), we have
only one free parameter, the radius R of the the sectional
curvature of the projective Hilbert state space.

3.Terms which arise in the Lagrangian (6.20) under
geodesic variation depend on {\it relative amplitudes}
of the
scalar field and they are connected  with
quantum transitions between different modes of this field.
One can relate these terms to some gauge ``surrounding field''.

4.Local ``non-abelian'' gauge transformations in the 
tangent bundle
contain {\it true covariant derivatives relative to the
Fubini-Study metric} in $CP(\infty)$ or, 
in some approximation, in $CP(N-1)$.  

5.The new Lagrangian (6.20) looks like the Lagrangian of 
the classical field but it should be treated as a 
Lagrangian of a quantum field as
it is obtained from the quantum projective state space.

6.The new Lagrangian gives the ``Higgs mechanism'' due to
form of the connection (4.1) in $CP(N-1)$ in an
absolutely natural way. 

The equation of motion of the self-interacting field configuration (droplet) may be obtained from variation
of the Lagrangian (6.20) relative to the variation of
$|\Psi>$. One has a generalized Klein-Gordon
equation
\begin{large}
\begin{eqnarray}
\frac{\partial^2(\Psi+\Delta\Psi)^*}{\partial(x^0)^2}-
\nabla^2(\Psi+\Delta\Psi)^*+\alpha^2(\Psi^*+\Delta\Psi^*+
\Psi^*\frac{\delta\Delta\Psi}{\delta \Psi})\cr
+\frac{\partial(\frac{\partial\Psi^*}{\partial x^0}
\frac{\delta\frac{\partial\Delta\Psi}{\partial x^0}}
{\delta\frac{\partial\Psi}{\partial x^0}})}{\partial x^0}
-\frac{\partial(\frac{\partial\Psi^*}{\partial x^1}
\frac{\delta\frac{\partial\Delta\Psi}{\partial x^1}}
{\delta\frac{\partial\Psi}{\partial x^1}})}{\partial x^1}
-\frac{\partial(\frac{\partial\Psi^*}{\partial x^2}
\frac{\delta\frac{\partial\Delta\Psi}{\partial x^2}}
{\delta\frac{\partial\Psi}{\partial x^2}})}{\partial x^2}
-\frac{\partial(\frac{\partial\Psi^*}{\partial x^3}
\frac{\delta\frac{\partial\Delta\Psi}{\partial x^3}}
{\delta\frac{\partial\Psi}{\partial x^3}})}{\partial x^3}
=0.\cr 
\end{eqnarray}
\end{large}
This equation is local in $CP(N-1)$ because this is connected
with the local topological vacuum $\Psi^0 \neq 0$. An analogous equation may be written in every sheet of the atlas. 
One can represent $\Delta\Psi$ for enough small $\tau$
with following Fourier coefficients
\begin{equation}
\Delta \Psi^i=- \frac{g\Psi^0\tau^2}
{\sqrt{1+\frac{|\Psi^0|^2}{R^2}}}\Gamma^i_{km}\xi^k \Psi^m. \end{equation}
It is easily to see that if the radius $R$ of the sectional curvature $1/R^2$ of the projective Hilbert space goes to infinity, one obtains the ordinary Klein-Gordon equation. In the general case the curvature of the projective state space influences the wave dispersion 
of a nonliner solution of the equation (6.23). This may be
treated as a base of the experimental testing of a quantum
nonlinearity in the sense, which was mentioned above.
This topic will be investigated in the near future.

The equation (6.23) presumably possesses localizable solutions like solitons then one can treat such 
solutions as {\it
primordial nonlocal elements of quantum theory} instead
of ``material points''. Furthermore, equation (6.23)
effectively describes propagation of the self-interacting
scalar field in a curved ``dynamical spacetime''.
That is the curvature of the projective state space may
be related with the curvature of the Einstein's ``dynamical spacetime'', i.e. with gravity. This problem requires
developments which will be discussed elsewhere.

\section{Discussion}
I believe that deep changes in the foundation of quantum
theory which have been proposed above should have an
influence on plans of some experiments.
I am {\it almost sure} that experiments like STEP (this experiment had been mentioned, for example, in Ref. [28]) {\it must} show
violations of the weak equivalence principle. However, it
does not mean that there exists ``fifth force'' or 
something of this kind. At so deep a quantum level as 
discussed in Ref. [28] such notions as, for example, ``acceleration'' in spacetime do not have a clear 
physical meaning. We should
analyse the dynamics of quantum states in the state space.
But in the classical limit where ``material point'', ``spacetime'' etc., have ordinary sense, the equivalence
principle of Einstein will be absolutely ``robust''.

ACKNOWLEDGMENTS

I thank Yuval Ne'eman and Larry Horwitz  for numerous useful
discussions and Yakir Aharonov for attention to this work.
This research was supported in part by grant PHY-9307708
of the National Science Foundation, and by grant of the
Ministry of Absorption of Israel.


\begin{thebibliography}{9}
\bibitem{1}
P.Dirac, Proc.Roy.Soc., A \bf 114\rm, 243 (1927).
\bibitem{2}
P.Dirac, Proc.Roy.Soc., A \bf 117\rm, 610 (1928).       
\bibitem{3} 
E.Schr\"odinger, Naturwissenschaften,\bf 14\rm, 664 (1926). 
\bibitem{4.}
E.Schr\"odinger, Ann. Physik, \bf 79\rm, 361 (1926).
\bibitem{5}
E.Schr\"odinger, Ann. Physik, \bf 79\rm, 489 (1926). 
\bibitem{6}
P.Leifer, \it Dynamics of the spin coherent state, Ph.D. Thesis,
Institute for Low Temperature Physics and Engineering of the
UkrSSR Academy of Science, \rm Kharkov, 1990.
\bibitem{7}
P.Leifer, On a Nonlinear Nonperturbative Modification
of Quantum Mechanics,\\ Preprint TAUP 2262-95.  
\bibitem{8}
L.Horwitz and P.Leifer, The Semiclassical Limit
from the Geometry of the Projective State Space, 
Preprint TAUP 2277-95 (submitted for publication).
\bibitem{9}
L.P.Horwitz and P.Leifer, Stable Wave Packet as a Model 
of an Elementary Particle,  Preprint TAUP 2302-95. 
\bibitem{10}
Y.Ne'eman,Phys.Lett.A \bf 186\rm, 5 (1994).
\bibitem{11}
P.Leifer, Nonlinear Modification
of Quantum Mechanics, \\ Preprint TAUP 2337-96 (submitted 
for publication)
\bibitem{12}
D.Bohm,{\it Causality and Chance in Modern Physics},
(London, 1957).
\bibitem{13}
A.M. Perelomov, {\it Generalized Coherent States and Their Applications},\\ Springer-Verlag, 1986.
\bibitem{14}
S.Kobayashi and K.Nomizu,{\it Foundations of Differntial Geometry,vol.II},\\
(Interscience Publishers, New York-London-Sydney, 1969).
\bibitem{15}
R.Cirelli, A. Mania', L. Pizzocchero, Int.Mod.Phys.,\bf 6 \rm No.12, 2133 (1991).
\bibitem{16}
S. Weinberg, Ann. Phys. (N.Y.) \bf 194\rm, 336 (1989).
\bibitem{17}
S. Weinberg, Phys.Rev.Lett. \bf 62\rm, 485 (1989).
\bibitem{18}
A.Shapere and F.Wilczek, J.Fluid Mech. \bf 198\rm, 557 (1989).
\bibitem{19}
M.Born and E.Wolf,{\it Principles of Optics},\\
(Pergamon Press, Oxford, London, New-York, 1964).
\bibitem{20}
L.de-Broglie, Compes Rend.,\bf 180\rm, 498 (1925).
\bibitem{21}
Janke-Emde-L\"osch,{\it Tafeln H\"oherer Funktionen}, Stuttgart (1960). 
\bibitem{22}
Yu.P.Rubakov and B.Saha, Found.Phys. \bf 25\rm, No.12,
1723 (1995).
\bibitem{23}
L.P.Horwitz, Found.Phys. \bf 22\rm, 491 (1992).
\bibitem{24}
M.M.Lam and C.Dewdney, Found.Phys. \bf 24\rm, No.1,
3 (1994).
\bibitem{25}
M.-T.Jeakel and S.Reynaud, Phys.Rev.Lett.\bf 76\rm, 
2407 (1996).
\bibitem{26}  
K.R.W. Jones, Ann.Phys. \bf 233\rm, 295 (1994).
\bibitem{27}  
Y. Ne'eman, Found.Phys. \bf 16\rm, 361 (1986).
\bibitem{28}
E.Fischbach et al., Phys.Rev.D \bf 52\rm, 5417 (1995).
\end{thebibliography}
\end{document}